\documentclass[runningheads]{llncs}
\usepackage[T1]{fontenc}
\usepackage{graphicx}

\usepackage{booktabs}   %
\usepackage{makecell}
\usepackage{dcolumn}    %
\usepackage{caption}
\usepackage{cite}
\usepackage{multirow}
\usepackage{amsmath}
\usepackage{amssymb}
\usepackage{placeins}   %

\begin{document}
\title{Beyond MSE: Rician Likelihood Denoising for Self-Supervised Cardiac $T2$ and $T1\rho$ MRI}
\titlerunning{Rician Likelihood Denoising}

\author{Nicholas A. Jacobs\inst{1}\orcidID{0009-0008-3277-260X} \and
Jason Mendes\inst{1}\orcidID{0000-0001-8146-535X} \and
Ravi Ranjan\inst{1}\orcidID{0000-0002-3321-2435} \and
Edward DiBella\inst{1}\orcidID{0000-0001-9196-3731}$^{,\dagger}$ \and
Shireen Elhabian\inst{1}\orcidID{0000-0002-7394-557X}$^{,\dagger}$}

\authorrunning{N. Jacobs et al.}

\let\svthefootnote\thefootnote
\let\thefootnote\relax
\footnotetext{$^{\dagger}$Senior authors}
\footnotetext{This preprint has not undergone peer review (when applicable) or any post-submission improvements or corrections. The Version of Record of this contribution is published at STACOM 2026.}
\let\thefootnote\svthefootnote

\institute{$^{1}$University of Utah, Salt Lake City, UT 84112, USA\\
\email{nicholas.jacobs@utah.edu}}

\maketitle

\begin{abstract}

Magnetic resonance imaging involves an inherent trade-off among spatial resolution, acquisition time, and noise. This trade-off contributes to long scan times and high cost. Deep learning has improved image denoising, but cardiac MRI remains difficult because high-resolution, rapid acquisitions generally lack corresponding low-noise ground truth. Self-supervised denoising offers a potential solution by learning from noisy image pairs or even single noisy acquisitions. However, we show that Noise2Void-style blind-spot denoising, which uses a mean squared error (MSE) loss and assumes zero-mean, independent and identically distributed (i.i.d.) noise, is poorly suited to MR magnitude images. When applied to short-axis $T2$-weighted and $T1\rho$-weighted cardiac MRI with synthetic Rician noise, it produces biased denoised images and biased parametric maps of $T2$ and $T1\rho$. To address this limitation, we formulate self-supervised denoising as maximum likelihood estimation under a known Rician noise model. This yields unbiased denoisers that are competitive with supervised baselines.


\end{abstract}

\section{Introduction}
Parametric mapping is common across many clinical MRI workloads, including cardiac, neural, and musculoskeletal imaging. It images tissue-specific magnetic properties that contrast-weighted images alone cannot show. Multiple contrast-weighted images are acquired and fit to an empirical model to estimate constants describing magnetization decay or growth. For example, $T1\rho$ is an emerging contrast-agent-free mechanism for myocardial tissue characterization \cite{han2014myocardiacdiseaseT1p, bustin2023mrcardiacT1pmapping}, but like other contrasts, it is limited by noise.

When acquiring MR images, there is an inherent trade-off among image resolution, acquisition time, and signal-to-noise ratio (SNR). This trade-off is especially pronounced in cardiac imaging, where acquisition time is dictated by the patient's cardiac cycle, forcing a choice between resolution and SNR.

Both existing approaches to denoising fall short. Supervised methods require paired clean and noisy data, which is unrealistic as high-resolution, low-noise cardiac images can be rare and impractical to acquire. The same holds for low-field scanners, which use weaker magnets to reduce cost and improve implant compatibility. Signal decreases proportionally to field strength, so these more accessible scanners operate in exactly the low-SNR regime where paired clean data is least available.

Self-supervised methods need no paired data, making them more applicable, but on magnitude MRI they produce biased results. Standard clinical magnitude images are not corrupted by additive Gaussian noise, but instead asymmetric Rician noise whose expectation lies above the true signal. A Gaussian approximation holds only at high SNR and fails at the low SNR where denoising is needed most. Because a regression objective recovers a summary statistic of the noisy signal (the conditional mean for MSE, the median for MAE) rather than the clean image, this bias cannot be removed by standard self-supervised training. Even MRI-specific self-supervised methods such as Noise2Contrast \cite{Wagner_Noise2Contrast_2022} train on MSE and inherit it.

We instead frame self-supervised denoising as maximum likelihood estimation under the known Rician noise model. Neither framing blind-spot denoising as maximum likelihood estimation nor using a Rician likelihood to train self-supervised MRI models is new on its own, but their combination is. Probabilistic Noise2Void (PN2V) \cite{Krull_PN2V_2020} generalizes blind-spot denoising with a per-pixel likelihood objective under an arbitrary noise model. Instead of analytically representing the conditional likelihood of a noisy observation given a clean signal, they represent it empirically with a histogram, built from paired clean and noisy calibration data. Our method instead uses an analytic Rician likelihood parametrized by a single scalar noise level, requiring no paired calibration data. This also allows us to denoise existing datasets where calibration data was not acquired, as the scalar noise level can be estimated retrospectively. Parker et al. \cite{Parker_RicianLikelihoodLoss_2025} show that a negative-log-Rician (NLR) loss removes the signal-dependent bias in self-supervised quantitative MRI. However, Parker et al. \cite{Parker_RicianLikelihoodLoss_2025} operate voxel-wise on magnitude images to directly predict diffusion parameters (ADC, IVIM) without any spatial prior. Neither applies an analytic Rician likelihood within a spatial blind-spot image denoiser. We do exactly this, using a known scalar noise level, and show it removes the bias that regression-based blind-spot training introduces in cardiac $T2$ and $T1\rho$ mapping.

We make three contributions. First, we show that regression-based blind-spot denoising in the style of Noise2Void \cite{Krull_Noise2Void_2019}, which assumes zero-mean i.i.d. noise, produces both biased denoised images and significantly biased parametric maps on Rician-corrupted cardiac MRI data. We derive this bias analytically for the MSE loss and show empirically that MAE suffers the same bias. Second, we apply an analytic Rician negative-log-likelihood loss within a spatial blind-spot image denoiser, combining the likelihood-based blind-spot training of PN2V \cite{Krull_PN2V_2020} with the NLR loss of Parker et al. \cite{Parker_RicianLikelihoodLoss_2025}, but in image space with a known scalar noise level, a setting neither addresses. Third, we demonstrate that self-supervised Rician Likelihood Denoising yields unbiased weighted-image denoising competitive with fully supervised baselines, with the bias reduction carrying through to the downstream $T2$ and $T1\rho$ maps, without paired data.

\section{Related Work}

Most denoising methods, such as DnCNN \cite{Zhang_DnCNN_2017}, treat denoising as a supervised task requiring paired clean and noisy images. Noise2Noise \cite{Lehtinen_Noise2Noise_2018} showed clean targets are unnecessary. A network trained to map one noisy realization to another converges to the clean signal, but still requires two aligned noisy acquisitions.

Noise2Void \cite{Krull_Noise2Void_2019} removes the paired-image requirement by framing denoising as masked prediction from surrounding context, operating on single noisy images. As we show, its implicit zero-mean noise assumption makes it biased on Rician magnitude data. Probabilistic Noise2Void \cite{Krull_PN2V_2020} generalizes this to an arbitrary noise model, but requires calibration data.

Noise2Contrast \cite{Wagner_Noise2Contrast_2022} extends blind-spot denoising by leveraging the multiple contrasts acquired in a single MRI session through joint domain translation. Unlike Rician Likelihood Denoising, it requires paired, registered contrasts, which are difficult to obtain under cardiac motion. It also trains with an MSE objective and therefore inherits the Rician bias our method removes. The closest work to ours is Parker et al. \cite{Parker_RicianLikelihoodLoss_2025}, whose NLR loss we adopt; the two differ as detailed in Section~1.

\section{Methods}

\textbf{Loss Theory.} For a blind-spot network such as Noise2Void \cite{Krull_Noise2Void_2019}, let $\tilde{x}_i$ be a noisy pixel with clean signal $s_i$, and $\tilde{x}_{\mathrm{RF}(i)}$ the receptive field with the center masked out. The network predicts $s_i$ from $\tilde{x}_{\mathrm{RF}(i)}$, yielding $\hat{s}_i$. As $\tilde{x}_i$ is excluded from the input, the network cannot learn the identity function. Noise2Void and similar work minimize the mean squared error $\mathrm{MSE}(\hat{s}_i, \tilde{x}_i)$ under the assumption of pixel-wise independent, zero-mean Gaussian noise, $\tilde{x}_i = s_i + n_i$ with $n_i \sim \mathcal{N}(0, \sigma^2)$. Minimizing it, the network learns the conditional expectation of its target given the context, which decomposes by linearity of expectation.

\begin{equation}
    \hat{s}_i \;\longrightarrow\;
    \mathbb{E}[\tilde{x}_i \mid \tilde{x}_{\mathrm{RF}(i)}]
    = \mathbb{E}[s_i \mid \tilde{x}_{\mathrm{RF}(i)}]
    + \mathbb{E}[n_i \mid \tilde{x}_{\mathrm{RF}(i)}]
\end{equation}

The noise term vanishes in two steps: $n_i$ is independent of the context $\tilde{x}_{\mathrm{RF}(i)}$, as it does not contain any of the spatially correlated signal and is independent of the surrounding noise. Therefore, $\mathbb{E}[n_i \mid \tilde{x}_{\mathrm{RF}(i)}] = \mathbb{E}[n_i]$, which equals $0$ by the zero-mean assumption. This leaves $\mathbb{E}[\tilde{x}_i \mid \tilde{x}_{\mathrm{RF}(i)}] = \mathbb{E}[s_i \mid \tilde{x}_{\mathrm{RF}(i)}]$, the minimum mean squared error estimate of the clean signal, approximating $s_i$ up to the information lost by masking the center.

MRI noise is Gaussian in the complex domain. It is i.i.d. across pixels, but the nonlinear transform to magnitude images creates Rician noise: the magnitude observations are conditionally independent given the signal, though correlated across pixels without conditioning, since the signal is correlated. The following equation describes the noisy pixel $\tilde{x}_i$ and how we generate synthetic noise.

\begin{equation}
    \tilde{x}_i = \sqrt{ \left( s_i + n_i^{\mathrm{Re}} \right)^2 + \left( n_i^{\mathrm{Im}} \right)^2 },
    \quad n_i^{\mathrm{Re}}, n_i^{\mathrm{Im}} \sim \mathcal{N}(0, \sigma^2)
    \label{eq:rician_distribution}
\end{equation}

Under this noise model the network again converges to the conditional expectation of the noisy pixel,

\begin{equation}
    \hat{s}_i \;\longrightarrow\;
    \mathbb{E}[\tilde{x}_i \mid \tilde{x}_{\mathrm{RF}(i)}]
    \label{eq:rician_convergence}
\end{equation}

Applying the law of total expectation \cite{Wackerly_Mendenhall_Scheaffer_2008}, in its conditional form $\mathbb{E}[X \mid Z] = \mathbb{E}[\mathbb{E}[X \mid Y,Z] \mid Z]$, to Equation~\ref{eq:rician_convergence} gives

\begin{equation}
    \mathbb{E}[\tilde{x}_i \mid \tilde{x}_{\mathrm{RF}(i)}]
    = \mathbb{E}[ \mathbb{E}[\tilde{x}_i \mid s_i, \tilde{x}_{\mathrm{RF}(i)}] \mid  \tilde{x}_{\mathrm{RF}(i)}]
    \label{eq:rician_expectation_full}
\end{equation}

Because the noise is pixel-wise independent conditioned on the signal, once $s_i$ is known $\tilde{x}_{\mathrm{RF}(i)}$ provides no new information; therefore $\mathbb{E}[\tilde{x}_i \mid s_i, \tilde{x}_{\mathrm{RF}(i)}] = \mathbb{E}[\tilde{x}_i \mid s_i]$, and Equation~\ref{eq:rician_expectation_full} simplifies to

\begin{equation}
    \mathbb{E}[ \mathbb{E}[\tilde{x}_i \mid s_i]  \mid  \tilde{x}_{\mathrm{RF}(i)}]
\end{equation}

As the Rice distribution is asymmetric, $\mathbb{E}[\tilde{x}_i] \neq s_i$, so a network trained with standard Noise2Void-style \cite{Krull_Noise2Void_2019} masked denoising cannot faithfully reconstruct the underlying clean, bias-free signal $s_i$. This motivates treating the problem as maximum likelihood estimation under the true Rician distribution. We adopt the negative log Rician likelihood (NLR) loss that Parker et al. \cite{Parker_RicianLikelihoodLoss_2025} derive via MLE, but apply it to blind-spot image denoising: our networks operate in magnitude image space, whereas Parker et al. predict diffusion parameters voxel-wise with no spatial prior.

We adopt the Rician negative log-likelihood loss formulation of Parker et al. \cite{Parker_RicianLikelihoodLoss_2025}, implemented independently.\footnote{The first log term follows directly from the log of the Rician pdf; note it differs from the expression printed in \cite{Parker_RicianLikelihoodLoss_2025}, which carries an extra factor of two in the denominator.} The NLR loss averaged over all pixel indices $i$ is shown below, where $I_0^e$ is the exponentially scaled modified Bessel function of the first kind with order zero, available in PyTorch as \texttt{torch.special.i0e}. The loss follows from taking the log of the Rician pdf and using the numerical stability trick of \cite{Parker_RicianLikelihoodLoss_2025}, computing $\log (I_0(z))$ as $\log(I_0^e(z)) + z$ rather than directly.

\begin{equation}
    L_{NLR} = - \frac{1}{N} \sum_{i=1}^{N} \left[
        \log\!\left( \frac{\tilde{x}_i}{\sigma^2} \right)
        - \frac{\tilde{x}_i^2 + s_i^2}{2\sigma^2}
        + \log I_0^e\!\left( \frac{\tilde{x}_i s_i}{\sigma^2} \right)
        + \frac{\tilde{x}_i s_i}{\sigma^2}
    \right]
\end{equation}

\noindent\textbf{Synthetic Noise.} Because our method trains without labels but clean references are still required to compute quantitative metrics, we add synthetic Rician noise to our data following Equation~\ref{eq:rician_distribution}. A single noise realization is written to disk for each data split, so the self-supervised models see only one realization, consistent with denoising raw scanner acquisitions to maximize SNR. The sole exception is the supervised training set, where noise realizations are generated on the fly and the model sees a different realization each epoch.

\noindent\textbf{Architectures.} We train three architectures under each form of supervision: fully supervised, self-supervised Noise2Void-style \cite{Krull_Noise2Void_2019}, and self-supervised Rician likelihood denoising. Both supervised and Noise2Void variants are trained with an L1 loss rather than MSE, as L1 is more robust to outliers and tends to yield higher perceptual quality \cite{zhao2016lossfunctionsforrestoration}; this choice is consistent with prior restoration work \cite{Zamir_Restormer_2022, liang2021swinirimagerestorationusing}. All models are multi-contrast: a single network handles both $T2$ and $T1\rho$. Models are trained on $32\times32$ patches, matching the local nature of denoising and prior blind-spot work \cite{Krull_Noise2Void_2019, Lehtinen_Noise2Noise_2018}. We additionally train one fully supervised UNet on full-sized images.

For a convolutional baseline we use a UNet \cite{Ronneberger_UNet_2015} with a ResNet-50 encoder \cite{He_Resnset_2016}. We also train two dense vision transformers \cite{dosovitskiy2021imageworth16x16words} with pixel-wise attention, which matches the pixel-wise granularity of the masking and the conditionally independent noise, and lets the model attend to small features consistent with the locality assumption imposed by patching. Each transformer has three layers with three $24$-dimensional heads and a feed-forward hidden dimension of $72$. We train ViTs with both learned position embeddings and rotary position embeddings (RoPE) \cite{su2023roformerenhancedtransformerrotary}. Inputs are tokenized by linearly projecting the $5\times5$ neighborhood of each pixel to the embedding dimension. We use softplus output ($\beta = 5$) for likelihood models, which require strictly positive predictions, and a linear output with post-hoc ReLU for regression models.

\noindent\textbf{Masking and Inference.} During training we select 16 pixels per patch, sampled uniformly with replacement. For each selected pixel, we replace its value with that of a neighbor drawn uniformly from the surrounding $5\times5$ window, which includes the pixel itself. The network is not told which pixels were masked, and the loss is computed only at the masked indices.

At evaluation no pixels are perturbed. Images are processed in overlapping patches with 25\% overlap, and predictions in overlapping regions are combined by Gaussian-weighted averaging, with the Gaussian standard deviation set to one-eighth of the patch side length.

\noindent\textbf{Metrics.} All metrics are reported over the manually segmented LV ROI. To separate systematic from random error, we decompose the mean squared error via the identity $\mathrm{MSE}(X,Y) = \mathrm{var}(X-Y) + \mathbb{E}[X-Y]^2$ \cite{Murphy_2012}. Per-image bias is the mean intensity difference over the ROI; we aggregate by averaging the absolute per-image biases, so that biases of opposite sign do not cancel and understate the error. Per-image variance is the population variance of the per-pixel difference from ground truth. Variance and RMSE, both non-negative, are reported as dataset means.

\section{Experiments}

\noindent\textbf{Dataset.} We use an in-house dataset of short-axis 2D canine cardiac MRI from nine subjects, comprising 302 series and 1,271 weighted images, split into disjoint training, validation, and test sets by subject. A preclinical dataset was used for data availability. Because canines have faster heartbeats than humans, there is less acquisition time and more noise, making this a harder problem than human cardiac imaging. All scans were acquired on a 3\,T Siemens Prisma scanner using a gradient-echo readout with $1.6\times1.6\times8$\,mm resolution, a $12^\circ$ flip angle, and GRAPPA parallel imaging (acceleration factor 2, 36 reference lines). Networks are trained only on the training set; the validation set is used for hyperparameter tuning, and all results are reported on the test set. The dataset primarily contains $T2$- and $T1\rho$-weighted images, plus 429 combined-weighting images added to the training set to enlarge it; these are excluded from validation and test to match the clinical distribution. Low-quality series, for example, those with significant artifact, were manually identified and removed from validation and test, as such acquisitions would be repeated in practice. Series contain three to twelve weighted images, most often three or four. In the test set, $T2$ series have three images at echo times 0, 30, and 55~ms, and $T1\rho$ series have four images acquired with a 500~Hz spin-lock frequency at spin-lock times 0, 30, 60, and 60~ms.

\begin{table}[tbp]
\centering
\caption{Dataset composition by split. Series counts, with weighted-image counts in parentheses.}
\label{tab:dataset-composition}
\setlength{\tabcolsep}{6pt}
\begin{tabular}{lrrrr r}
\toprule
      & Subjects & T2 & T1$\rho$ & \makecell{Combined\\T2/T1$\rho$} & Total \\
\midrule
Train & 5 & 69 (225) & 99 (460)  & 86 (429) & 254 (1,114) \\
Validation   & 2 & 11 (33)  & 13 (43)   & --       & 24 (76)    \\
Test  & 2 & 15 (45)  & 9 (36)    & --       & 24 (81)    \\
\midrule
Total & 9 & 95 (303) & 121 (539) & 86 (429) & 302 (1,271) \\
\bottomrule
\end{tabular}
\end{table}

\noindent\textbf{Preprocessing.} Images were reconstructed and registered online with standard Siemens algorithms, at either $384\times252$ or $192\times126$ pixels; the smaller were bilinearly upsampled to $384\times252$. Intensities were normalized per series by the 99th percentile, which is more robust than min/max scaling where a single noisy pixel can distort the result. Images were then symmetrically zero-padded to $384\times256$ for divisibility by 32 for the UNet. These serve as the clean ground truth against which all metrics are reported.

\noindent\textbf{Experimental Setup.} All models were trained for 200 epochs in fp16 mixed precision. Patch-based models used 1024 patches per batch, with 96 random $32\times32$ crops taken per image each epoch and shuffled across batches. We set this to the number of non-overlapping patches in a $384\times256$ image, so that each epoch corresponds to roughly one full pass over the data. The full-image supervised UNet used 64 images per batch.

\noindent\textbf{Results.} Table~\ref{tab:weighted-results} reports denoising performance over the LV for weighted images in the test set. The variance column is omitted, as it was approximately 0.0004 and method-invariant. With spatial variance constant, the excess RMSE of Noise2Void \cite{Krull_Noise2Void_2019} relative to the supervised models, higher for both $T2$ and $T1\rho$, is almost entirely bias: a minor increase in RMSE tracks a significant increase in bias. Switching from Noise2Void \cite{Krull_Noise2Void_2019} to a Rician likelihood reduces this bias by up to a factor of 8.4, recovering supervised-level performance. Although omitted for brevity, the full-image UNet achieved higher SSIM than the patch-based UNet (0.88 vs. 0.85) despite worse LV accuracy, suggesting that global context aids perceptual quality at a cost to quantitative performance.

\begin{table}[htb]
\centering
\caption{Denoising performance for weighted images on the LV ROI.}
\label{tab:weighted-results}
\setlength{\tabcolsep}{6pt}
\begin{tabular}{ll rr rr}
\toprule
 & & \multicolumn{2}{c}{T2} & \multicolumn{2}{c}{T1$\rho$} \\
\cmidrule(lr){3-4} \cmidrule(lr){5-6}
 & & Bias & RMSE & Bias & RMSE \\
\midrule
\multirow{4}{*}{Supervised}
      & UNet (images)  & 0.0048 & 0.021 & 0.0029 & 0.023 \\
      & UNet (patches) & 0.0033 & 0.018 & 0.0027 & 0.020 \\
      & ViT            & 0.0066 & 0.021 & 0.0061 & 0.023 \\
      & ViT with RoPE  & 0.0044 & 0.021 & 0.0035 & 0.022 \\
\midrule
\multirow{3}{*}{Noise2Void}
      & UNet (patches) & 0.0253 & 0.033 & 0.0255 & 0.034 \\
      & ViT            & 0.0226 & 0.032 & 0.0231 & 0.033 \\
      & ViT with RoPE  & 0.0253 & 0.034 & 0.0265 & 0.035 \\
\midrule
\multirow{3}{*}{\makecell[l]{Rician\\Likelihood\\Denoising}}
      & UNet (patches) & 0.0030 & 0.020 & 0.0028 & 0.021 \\
      & ViT            & 0.0042 & 0.023 & 0.0027 & 0.024 \\
      & ViT with RoPE  & 0.0054 & 0.024 & 0.0029 & 0.025 \\
\bottomrule
\end{tabular}
\end{table}

\begin{figure}[htb]
    \centering
    \includegraphics[width=0.99\linewidth]{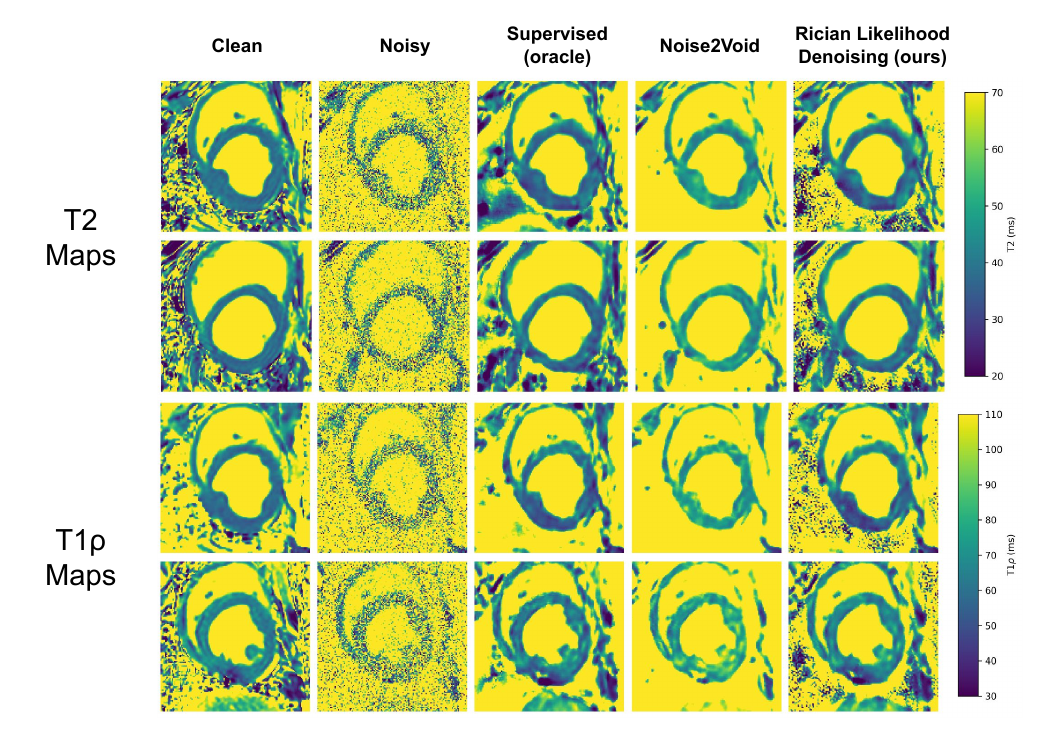}
    \caption{$T2$ and $T1\rho$ maps for two representative test series per contrast with results for UNet models. Noise2Void \cite{Krull_Noise2Void_2019} retains the Rician bias, brightening the myocardium, whereas Rician Likelihood denoising recovers reference values and matches the supervised results.}
    \label{fig:compare}
\end{figure}

Table~\ref{tab:map-results} reports denoising performance over the LV for parameter maps in the test set. Because parameter fitting is nonlinear, the spatially varying bias in the weighted images propagates into both the bias and variance of the resulting maps. On the $T1\rho$ maps, the Noise2Void \cite{Krull_Noise2Void_2019} models are biased by almost 13~ms on average, which the Rician likelihood loss reduces to 1.15~ms while also lowering variance. Example maps from the test sets are shown in Figure~\ref{fig:compare}.

\begin{table}[!b]
\centering
\caption{Denoising performance by supervision and architecture for parameter maps. Bias (ms), variance (ms$^2$), and RMSE (ms) are reported on the LV ROI.}
\label{tab:map-results}
\setlength{\tabcolsep}{6pt}
\begin{tabular}{ll rrr rrr}
\toprule
 & & \multicolumn{3}{c}{T2} & \multicolumn{3}{c}{T1$\rho$} \\
\cmidrule(lr){3-5} \cmidrule(lr){6-8}
 & & Bias & Variance & RMSE & Bias & Variance & RMSE \\
\midrule
\multirow{4}{*}{Supervised}
      & UNet (images)  & 1.34 & 20.8 & 4.33 & 1.20  & 64.1  & 8.10  \\
      & UNet (patches) & 0.60 & 11.9 & 3.43 & 0.93  & 46.1  & 6.84  \\
      & ViT            & 1.79 & 13.7 & 4.08 & 2.54  & 55.1  & 7.89  \\
      & ViT with RoPE  & 1.05 & 13.9 & 3.83 & 1.32  & 56.2  & 7.64  \\
\midrule
\multirow{3}{*}{Noise2Void}
      & UNet (patches) & 6.52 & 37.3 & 8.57 & 11.87 & 106.7 & 15.68 \\
      & ViT            & 6.39 & 50.3 & 9.03 & 12.31 & 160.0 & 17.21 \\
      & ViT with RoPE  & 7.51 & 42.4 & 9.62 & 14.67 & 114.7 & 18.10 \\
\midrule
\multirow{3}{*}{\makecell[l]{Rician\\Likelihood\\Denoising}}
      & UNet (patches) & 0.60 & 13.6 & 3.67 & 1.19  & 53.3  & 7.39  \\
      & ViT            & 1.13 & 17.6 & 4.30 & 1.04  & 69.3  & 8.34  \\
      & ViT with RoPE  & 1.50 & 18.5 & 4.52 & 1.23  & 71.4  & 8.51  \\
\bottomrule
\end{tabular}
\end{table}

Model performance for $T1\rho$ maps is noticeably worse than $T2$. This likely stems from the noisier $T1\rho$ ground truth, which limits how accurately the metrics reflect denoising quality, and the less uniform preparation times in the $T1\rho$ dataset. Partially offsetting this, the $T1\rho$ test set has four weighted images per map versus three for $T2$, which better constrains the nonlinear fit at each pixel.

\section{Conclusion}

We showed that Noise2Void-style blind-spot denoising, which assumes zero-mean i.i.d. noise, is systematically biased on Rician magnitude MRI, and that framing denoising as maximum likelihood estimation under the true Rician model removes this bias. The resulting denoisers reduce parameter-map bias from nearly 13~ms to under 1.2~ms on $T1\rho$, matching fully supervised baselines while using only noisy images and a scalar noise level.

Our method is limited by its noise assumptions. Our experiments use synthetic Rician noise that satisfies the assumed model. However, on raw scanner data, modern reconstructions such as GRAPPA and compressed sensing produce spatially correlated noise that violates our pixel-wise-independence assumption. We also assume a known noise level. Although $\sigma$ can be estimated from background pixels, we have not characterized the method's sensitivity to a mismatch between the assumed and true noise level.

Several extensions remain for future work. The noise model could be generalized to spatially varying noise and to the non-central chi distribution, which better models parallel imaging. The locality and patching assumptions could also be relaxed: the image-level model achieved higher SSIM yet worse LV accuracy than the patch-based models, a global-versus-local trade-off worth exploring through sparse attention.

\begin{credits}
\subsubsection{\ackname} This work was supported by the National Heart, Lung, and Blood Institute of the National Institutes of Health under award NIH R01HL178117. The content is solely the responsibility of the authors and does not necessarily represent the official views of the National Institutes of Health.
\end{credits}

\FloatBarrier   %
\bibliographystyle{splncs04}
\bibliography{bibliography}

\begin{thebibliography}{10}
\providecommand{\url}[1]{\texttt{#1}}
\providecommand{\urlprefix}{URL }
\providecommand{\doi}[1]{https://doi.org/#1}

\bibitem{bustin2023mrcardiacT1pmapping}
Bustin, A., Witschey, W.R., Heeswijk, R.B.v., Cochet, H., Stuber, M.: Magnetic
  resonance myocardial t1$\rho$ mapping. Journal of Cardiovascular Magnetic
  Resonance  (2023)

\bibitem{dosovitskiy2021imageworth16x16words}
Dosovitskiy, A., Beyer, L., Kolesnikov, A., Weissenborn, D., Zhai, X.,
  Unterthiner, T., Dehghani, M., Minderer, M., Heigold, G., Gelly, S.,
  Uszkoreit, J., Houlsby, N.: An image is worth 16x16 words: Transformers for
  image recognition at scale. In: International Conference on Learning
  Representations (2021)

\bibitem{han2014myocardiacdiseaseT1p}
Han, Y., Liimatainen, T., Gorman, R.C., Witschey, W.R.T.: Assessing myocardial
  disease using t1$\rho$ mri. Current Cardiovascular Imaging Reports
  \textbf{7}(2), ~9248 (2014)

\bibitem{He_Resnset_2016}
He, K., Zhang, X., Ren, S., Sun, J.: Deep residual learning for image
  recognition. In: Proceedings of the IEEE Conference on Computer Vision and
  Pattern Recognition (CVPR) (June 2016)

\bibitem{Krull_Noise2Void_2019}
Krull, A., Buchholz, T.O., Jug, F.: Noise2void - learning denoising from single
  noisy images. In: Proceedings of the IEEE/CVF Conference on Computer Vision
  and Pattern Recognition (CVPR) (June 2019)

\bibitem{Krull_PN2V_2020}
Krull, A., Vičar, T., Prakash, M., Lalit, M., Jug, F.: Probabilistic
  noise2void: Unsupervised content-aware denoising. Frontiers in Computer
  Science  \textbf{2} (Feb 2020)

\bibitem{Lehtinen_Noise2Noise_2018}
Lehtinen, J., Munkberg, J., Hasselgren, J., Laine, S., Karras, T., Aittala, M.,
  Aila, T.: {N}oise2{N}oise: Learning image restoration without clean data. In:
  Dy, J., Krause, A. (eds.) Proceedings of the 35th International Conference on
  Machine Learning. Proceedings of Machine Learning Research, vol.~80, pp.
  2965--2974. PMLR (10--15 Jul 2018)

\bibitem{liang2021swinirimagerestorationusing}
Liang, J., Cao, J., Sun, G., Zhang, K., Van~Gool, L., Timofte, R.: Swinir:
  Image restoration using swin transformer. In: Proceedings of the IEEE/CVF
  International Conference on Computer Vision (ICCV) Workshops. pp. 1833--1844
  (October 2021)

\bibitem{Murphy_2012}
Murphy, K.P.: Machine learning: A probabilistic perspective. The MIT Press
  (2012)

\bibitem{Parker_RicianLikelihoodLoss_2025}
Parker, C.S., Schroder, A., Epstein, S.C., Cole, J., Alexander, D.C., Zhang,
  H.: Rician likelihood loss for quantitative mri with self-supervised deep
  learning. NMR in Biomedicine  \textbf{38}(10),  e70136 (2025)

\bibitem{Ronneberger_UNet_2015}
Ronneberger, O., Fischer, P., Brox, T.: U-net: Convolutional networks for
  biomedical image segmentation. In: Navab, N., Hornegger, J., Wells, W.M.,
  Frangi, A.F. (eds.) Medical Image Computing and Computer-Assisted
  Intervention -- MICCAI 2015. pp. 234--241. Springer International Publishing
  (2015)

\bibitem{su2023roformerenhancedtransformerrotary}
Su, J., Ahmed, M., Lu, Y., Pan, S., Bo, W., Liu, Y.: Roformer: Enhanced
  transformer with rotary position embedding. Neurocomputing  \textbf{568},
  127063 (2024)

\bibitem{Wackerly_Mendenhall_Scheaffer_2008}
Wackerly, D.D., Mendenhall, W., Scheaffer, R.L.: Mathematical statistics with
  Applications. Thomson Brooks/Cole; Cengage (2008)

\bibitem{Wagner_Noise2Contrast_2022}
Wagner, F., Thies, M., Pfaff, L., Maul, N., Pechmann, S., Gu, M., Utz, J.,
  Aust, O., Weidner, D., Neag, G., Uderhardt, S., Choi, J.H., Maier, A.:
  Noise2contrast: Multi-contrast fusion enables self-supervised tomographic
  image denoising. In: Information Processing in Medical Imaging: 28th
  International Conference, IPMI 2023, San Carlos de Bariloche, Argentina, June
  18-23, 2023, Proceedings. pp. 771--782. Springer-Verlag, Berlin, Heidelberg
  (2023)

\bibitem{Zamir_Restormer_2022}
Zamir, S.W., Arora, A., Khan, S., Hayat, M., Khan, F.S., Yang, M.H.: Restormer:
  Efficient transformer for high-resolution image restoration. In: Proceedings
  of the IEEE/CVF Conference on Computer Vision and Pattern Recognition (CVPR).
  pp. 5728--5739 (June 2022)

\bibitem{Zhang_DnCNN_2017}
Zhang, K., Zuo, W., Chen, Y., Meng, D., Zhang, L.: Beyond a gaussian denoiser:
  Residual learning of deep cnn for image denoising. IEEE Transactions on Image
  Processing  \textbf{26}(7),  3142--3155 (2017)

\bibitem{zhao2016lossfunctionsforrestoration}
Zhao, H., Gallo, O., Frosio, I., Kautz, J.: Loss functions for image
  restoration with neural networks. IEEE Transactions on computational imaging
  \textbf{3}(1),  47--57 (2016)

\end{thebibliography}

\end{document}